\documentstyle[floats,aps,epsf,prl]{revtex}

\begin{document}

\title {\bf Controlling adsorbate vibrational lifetimes using
superlattices
}
\author{Steven P. Lewis}
\address{Department of Physics and Astronomy and Center for 
Simulational Physics, \\
University of Georgia, Athens, GA 30602-2451.}
\author{Andrew M. Rappe}
\address{Department of Chemistry and Laboratory for Research 
on the Structure of Matter, \\ 
University of Pennsylvania, Philadelphia, PA 19104.}
\date{\today}
\maketitle

\begin{abstract} 
We propose using short-period superlattices as substrates to control 
the vibrational relaxation dynamics of adsorbate overlayers.  The mass 
modulation of superlattices creates both band gaps and large spectral 
enhancements in the phonon density of states.  These modifications
can dramatically alter the coupling between vibrational modes of the
adsorbate overlayer and the substrate lattice, thereby significantly
affecting the lifetime of adsorbate modes.
\end{abstract} 


\section{Introduction}

The lowest energy vibrational excitations of molecules adsorbed on metal
surfaces generally correspond to motion of the molecule as a whole about the 
adsorption bond, i.e., frustrated translations and rotations.  These modes 
play an important role in thermally activated processes at surfaces, such 
as catalysis and surface diffusion.\cite{Pers88}  It would, therefore, be 
highly desirable to be able to control the dynamical behavior of these 
vibrations.  We propose a way to dramatically alter the lifetime of 
low-frequency adsorbate modes by suitably modifying the composition of the 
substrate.

We recently developed a theory for the relaxation dynamics of adsorbate
vibrational modes whose frequency lies within the long-wavelength acoustic
part of the substrate phonon spectrum.\cite{Lew96,Lew98,Pyk98,Lew99}  
This frequency range is relevant, for example, for in-plane frustrated 
translations (FT) of molecules on metal surfaces.  We have shown that 
the relaxation of these modes is governed by resonant coupling to bulk 
substrate phonons; {\it i.e.,} the vibrating overlayer radiates its energy 
into resonant acoustic waves.\cite{others}  All other relaxation channels, 
such as substrate electron-hole pair excitation, are typically sub-dominant.
This result suggests that the lifetime of FT and other low-lying adsorbate 
modes can be significantly altered by changing the substrate phonon spectrum 
at the resonant frequency.

In this paper we explore the use of superlattices to achieve this goal.  For 
the last two decades, superlattices have received considerable attention by 
materials researchers because of their unusual and highly tunable properties.  
Superlattices consist of a periodic alternation of layers of two or more 
different materials.  For simplicity, this paper focuses on superlattices of 
two materials, $A$ and $B$.  The resulting periodic mass modulation causes 
frequency gaps to open in the phonon spectrum for modes propagating parallel 
to the stacking direction, $\hat{s}$.  The size and placement of the band gaps 
are determined by compositional parameters of the superlattice, such as the 
mass ratio of materials $A$ and $B$ and the modulation period.  The modes 
expelled from band-gap regions gather near van Hove singularities at the 
gap edges.  Thus, the superlattice mass modulation enhances the bulk phonon
spectrum at some frequencies while depleting it at others.  

We consider an ordered adsorbate overlayer on the surface of such a 
superlattice, with the surface normal aligned along $\hat{s}$.  A 
$\vec{k}_\| = 0$ FT excitation of the overlayer will only couple to
bulk modes propagating parallel to $\hat{s}$.  Thus, depending on the
superlattice compositional parameters, the FT frequency may fall in a
region of either depleted or enhanced phonon density of states (DOS), 
compared to the pure substrate.  In the former case, resonant phonon 
emission is eliminated as a FT decay channel, because there are no 
phonon states at the resonant frequency.  In the latter case, one might 
expect enhanced coupling because of the increased phase space.  However, 
before reaching that conclusion, it is necessary to consider the coupling 
strength to modes in the van Hove singularities.  In this work, we examine 
all of these issues using a model system corresponding to an ordered 
adsorbate overlayer on a face-centered cubic (fcc) superlattice.

Acoustic waves in superlattice substrates have been investigated previously.
\cite{Kue82a,Kue82b,Cam83,Car92,Bou96}  In particular, Camley {\it et al.} 
derived vibrational dispersion relations for both infinite and semi-infinite 
superlattices in the elastic-continuum limit ({\it i.e.,} both the solid as 
a whole and the $A$ and $B$ layers individually are treated as continuous 
media).\cite{Cam83}  They focused on superlattices with a modulation wavelength 
of 1000~\AA\ or more.  In this paper, we opt for an atomistic approach instead 
of an elastic-continuum model, because we are interested in superlattices of 
small enough modulation wavelength that variations by even a single atomic 
layer cause noticeable changes in the bulk phonon spectrum.  Nevertheless, 
the basic physical picture presented in Ref.~\onlinecite{Cam83} still applies.  

Figure 6 of Ref.~\onlinecite{Cam83} is a plot of $\omega$ vs.\ $k_\|$ for 
an infinite superlattice.  This plot shows that the mass modulation creates 
frequency gaps in the bulk vibrational spectrum not only at $k_\| = 0$, but 
also at finite $k_\|$.  Indeed, for any frequency, there will be regions
of the $k_\|$-plane for which there are no propagating acoustic modes and
other regions for which propagating acoustic modes exist.  Points on the 
boundary between these two types of regions are van Hove singularities.
This picture is very encouraging, since $k_\|$-resolved experimental surface 
techniques, such as inelastic helium-atom scattering,\cite{Hof96} could 
then be used to exhibit the whole range of behaviors predicted here for 
adsorbate vibrations coupled to a superlattice.  Camley {\it et al.,} also 
present $\omega$ vs.\ $k_\|$ for a semi-infinite superlattice (Fig.~8 of 
Ref.~\onlinecite{Cam83}).  They show that a band of surface waves appears 
in each bulk band-gap region.  

\section{Model}

We use lattice dynamics to study the vibrational coupling between the
adsorbate overlayer and the superlattice substrate.  In this approach,
the atoms are treated as point masses, and each atomic degree of freedom
of the system is coupled harmonically to all others.  The normal modes
of vibration are obtained by solving the generalized eigenvalue problem,
${\rm\bf K} \cdot x = -\omega^2 {\rm\bf M} \cdot x$, where the eigenvector 
$x$ resides in the 3$J$-dimensional configuration space of the $J$ atoms, 
with origin at the equilibrium configuration, the eigenvalue $\omega^2$ 
is the square of the normal-mode frequency, {\bf M} is the diagonal 
3$J$$\times$3$J$ matrix of atomic masses, and {\bf K} is the symmetric 
3$J$$\times$3$J$ matrix of harmonic force constants.  For metals and 
non-polar semiconductors and insulators, a cutoff can be defined so that 
only coupling constants above a given magnitude are included.  This makes 
the {\bf K} matrix sparse.

Our model substrate is an fcc superlattice with the stacking axis and 
the surface normal both parallel to the [001] crystallographic axis.  
Integers $N_A$ and $N_B$ represent the number of atomic (001) planes 
of materials $A$ and $B$, respectively, in each superlattice period 
($N = N_A + N_B$).  The overall thickness of the substrate in our 
calculations is 250$N$.  Our model overlayer consists of an ordered 
array of diatomic molecules forming a $c$(2$\times$2) pattern ({\it 
i.e.,} checkerboard pattern).  In-plane periodic boundary conditions
are used, making the model system of infinite extent parallel to the
surfaces and superlattice interfaces, but restricting our analysis to
$k_\| = 0$ modes.

For simplicity in this exploratory research, we make the approximation 
that the force constants coupling the substrate atoms are insensitive 
to the chemical species.  Therefore, varying the superlattice 
composition is achieved simply by altering the mass matrix {\bf M}.  
For real systems, the mass modulation accounts for the majority of the 
effect, with compositionally varying force constants contributing 
essentially no new physics.

The major conclusions of this research are insensitive to the 
specific details of the model system.  However, we study this 
particular structural model for two important reasons.  First, 
this structure is often experimentally realized for diatomic 
adsorbates on fcc-metal (001) substrates.  Second, this choice of 
model allows us to use the force constants we obtained from earlier 
density-functional theory calculations for the prototypical system 
of CO on Cu (001).\cite{Lew96,Lew99}  The vibrational frequencies 
computed in that study are in excellent quantitative agreement with 
experiment.  See Ref.~\onlinecite{Lew99} for numerical values of the 
force constants, as well as details of how they were computed and 
comparisons to experiment.  For definiteness and ease of comparison 
to our earlier work, we also choose the masses of CO and Cu for the 
adsorbate and $A$-atom masses, respectively. 

Given the model described above, there remain three parameters for 
specifying the system: the mass ratio $m_B/m_A$ of the two superlattice 
materials and the layer thicknesses, $N_A$ and $N_B$.  Experience shows 
that the mass ratio needs to be fairly large in order to create sizable 
phonon band gaps.  We have chosen a mass ratio of 3:1 ({\it e.g.,} Au 
and Cu have this mass ratio).  

The modulation period $N$ should be chosen so that the FT frequency is 
near or within a phonon band gap.  If the frequency of resonant phonons 
in the pure material corresponds to a wavelength spanning about $M$ 
atomic layers, then Bragg theory suggests that a modulation period of 
$N=M/2$ layers will open a band gap at or near the resonant frequency.  
For the model parameters corresponding to $c$(2$\times$2)-CO on Cu (001), 
our base system, this condition gives a modulation wavelength of about 
$N=8$ layers.  This value is to be contrasted with the elastic-continuum 
work of Ref.~\onlinecite{Cam83}, which considered a superlattice with 
1500~\AA\ modulation wavelength, corresponding to $N \approx 800$ layers.  
The resulting lowest band gap was 0.05 cm$^{-1}$ wide, centered at a 
frequency of 0.24 cm$^{-1}$.  These values are two orders of magnitude 
too small for our purposes, and we thus need to consider a superlattice 
modulation that is two orders of magnitude narrower.  The distribution 
of the $N = 8$ layers between materials $A$ and $B$ is left as a variable
parameter.

\section{Results}

\begin{figure}
\epsfysize=4.50in
\centerline{\epsfbox[18 194 592 718]{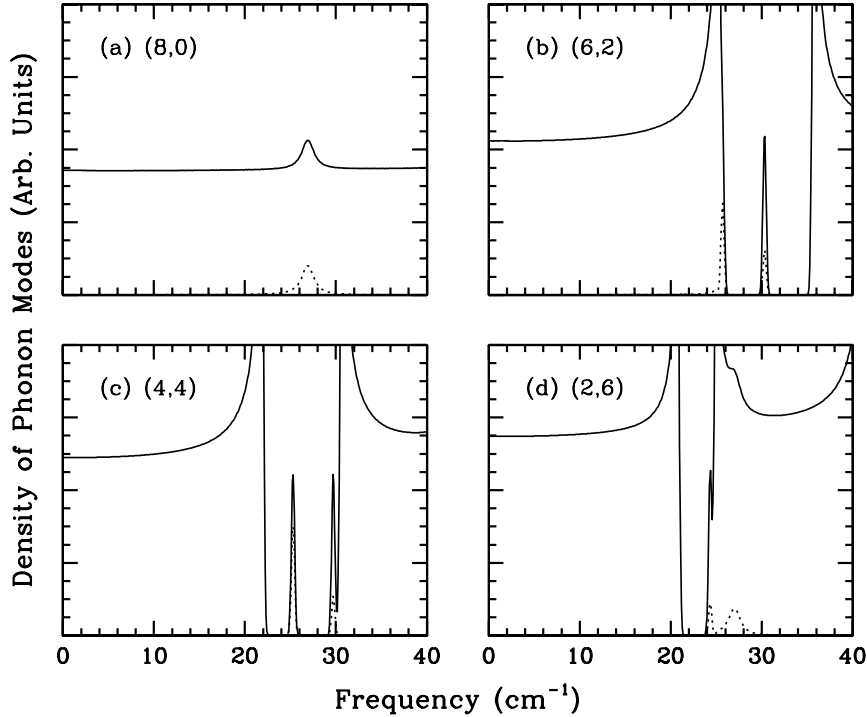}}
\caption{Vibrational spectrum for an adsorbed superlattice substrate with $N=8$. 
Each panel represents a different partitioning of the superlattice period into $A$
and $B$ layers.  In terms of the symbol ($N_A$,$N_B$) described in the text: (a) 
(8,0); (b) (6,2); (c) (4,4); (d) (2,6).  The solid curve in each panel gives the
full vibrational density of states, whereas the dashed curve gives the density of
states projected onto the frustrated-translational mode.
}
\label{dos}
\end{figure}

Figure \ref{dos}(a)-(d) shows the low-frequency portion of the 
vibrational DOS of the coupled adsorbate/substrate system for $N=8$ 
superlattices.  The dashed curve in each panel is the DOS projected 
onto the idealized FT coordinate.  Each panel corresponds to a 
different partitioning of the 8 layers of the superlattice period 
into $A$ and $B$ layers.  The symbol ($N_A$,$N_B$) in the figure 
panels specifies the partitioning.  The (8,0) panel (Fig.~\ref{dos}(a)) 
corresponds to a pure substrate of material $A$.  For the present 
choice of masses and force constants, this system is identical to our 
vibrational spectrum in Refs.~\onlinecite{Lew96} and online\cite{Lew99}.  
The broad peak centered at 27~cm$^{-1}$ is the FT resonance, whose 
natural line width corresponds to a lifetime of 3.0 ps.

As soon as mass modulation is introduced into the substrate, band gaps
open in the DOS near the FT frequency, as predicted.  In addition there 
are band gaps at higher frequencies corresponding to higher order Bragg 
reflections.  The band gaps are accompanied by the large spectral 
enhancement on either side of the gap characteristic of van Hove 
singularities.  Notice that isolated peaks often appear in the band gaps.  
These peaks are mixtures of the surface states described by Camley 
{\it et al.} in Ref.~\onlinecite{Cam83} with the adsorbate frustrated
translation.  They are truly $\delta$-functions; the width that appears 
in the figure results from Gaussian smoothing of the spectra.

For the (6,2) superlattice, the van Hove singularity below the band 
gap coincides in frequency with the FT mode (see Fig.~\ref{dos}(b)).  
Despite the large density of bulk phonon states at that frequency, 
the FT mode does not form a broad resonance.  Rather, it remains a 
narrow, $\delta$-function-like peak.  This suggests that the coupling 
strength of the FT to bulk modes in this peak is very weak.  We 
explore this issue further in the discussion below.

As the content of the more massive material ($B$) in the superlattice
increases, the band gap narrows and shifts to lower frequency.  These
changes reflect the overall contraction of much of the phonon spectrum 
due to the $1/\sqrt{M}$-like dependence of vibrational frequencies.
For the (4,4) superlattice, this effect causes the lower edge of the
band gap to come below the FT frequency, leaving the adsorbate mode 
in the gap with no phonons to couple to (see Fig.~\ref{dos}(c)).  
This feature is now an infinitely narrow $\delta$-function.  Resonant 
coupling to substrate phonons, therefore, contributes nothing to FT 
relaxation, leaving previously marginal processes, such as 
electron-phonon, multi-phonon, and anharmonic coupling, as the 
dominant decay channels.  

For the (2,6) superlattice, the narrowing and lowering of the gap
has caused the upper edge of the gap to come below the FT frequency,
leaving the FT mode once again well within the bulk phonon continuum.
The coupling to these bulk modes is strong enough to produce a broad
FT resonance peak, as in the case of the pure (8,0) substrate.

\section{Discussion}

\begin{figure}
\epsfysize=4.50in
\centerline{\epsfbox[18 194 592 718]{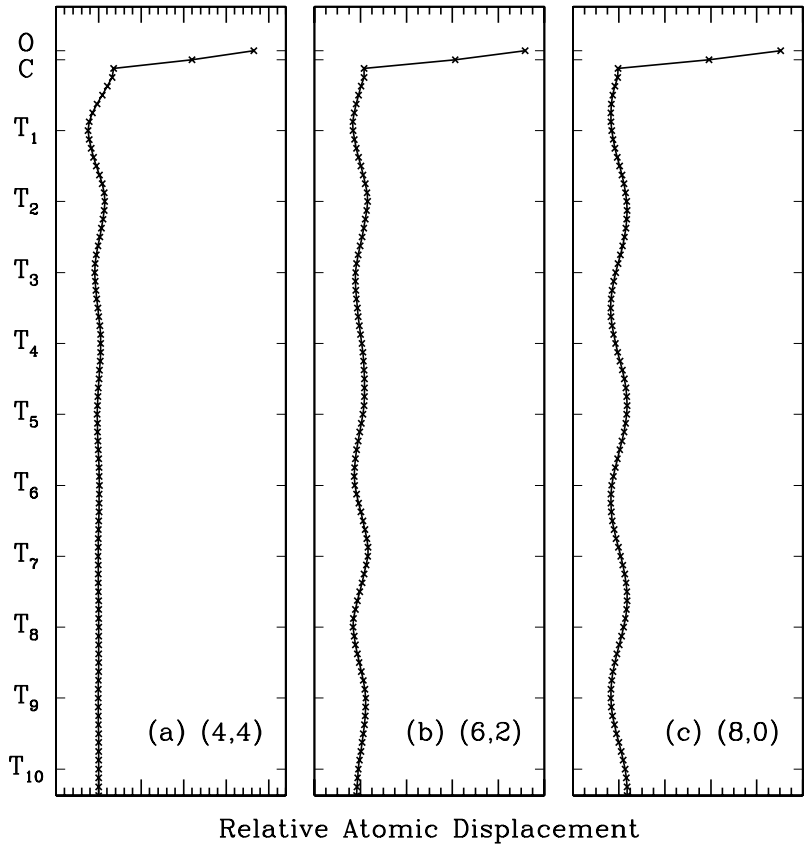}}
\caption{
Atomic displacement patterns for frustrated-translational normal modes.  In each 
panel, the vertical axis denotes the position of a layer of atoms relative to the 
surface, and the horizontal axis denotes the relative displacement of each layer 
away from equilibrium for the given normal mode.  The symbol $T_i$ denotes the 
interface between the $i$th and $(i+1)$th superlattice periods.  (a) Isolated FT 
mode in the (4,4) spectrum.  (b) Center of the FT resonance of the (6,2) spectrum.
(c) Center of the FT resonance of the (8,0) spectrum.
}
\label{eigvec}
\end{figure}

To gain a deeper understanding of the effect of varying superlattice 
composition on the FT dynamics, we examine, in Fig.~\ref{eigvec}, the
atomic displacement patterns of FT-related normal modes for the (4,4), 
(6,2), and (8,0) systems.  For the (6,2) and (8,0) systems, we display 
the modes at the center of the FT resonance peak.  In each panel, the
vertical axis denotes the position of a layer of atoms relative to 
the surface, and the horizontal axis denotes the relative displacement 
of each layer away from equilibrium for the given normal mode.  For
clarity only $\sim$80 atomic layers of the 2,000-layer substrate are 
shown.

Figure \ref{eigvec} highlights the characteristic delocalization 
that occurs when an isolated mode comes into resonance with a 
continuum of states.  The isolated FT mode for the (4,4) system, 
shown in Fig.~\ref{eigvec}(a), is exponentially localized at the 
surface.  The side-to-side ``wagging'' of the CO molecules dominates 
the displacement pattern, with the amplitude on substrate layers 
rapidly decaying as a function of penetration distance.  In contrast, 
the resonance modes pictured in Figs.~\ref{eigvec}(b) and (c) for the
(6,2) and (8,0) systems, respectively, are clearly combinations of 
adsorbate FT motion and substrate phonon motion.  These modes exhibit 
the large amplitude C-O wagging behavior characteristic of frustrated 
translations.  However, an appreciable amplitude of long-wavelength 
bulk-phonon motion persists throughout the substrate, so that the 
resonance states are not localized.  

Comparing panels (b) and (c) of Fig.~\ref{eigvec} shows that the 
substrate-phonon component of the mode has noticeably smaller amplitude 
for the (6,2) system than for the (8,0) system, indicating a smaller
coupling strength between the adsorbate FT motion and the substrate 
phonons.  We were able to infer the same result in the previous 
section by comparing both the DOS's and the FT-projected DOS's of the 
(6,2) and (8,0) systems.

\begin{figure}
\epsfysize=4.50in
\centerline{\epsfbox[18 194 592 718]{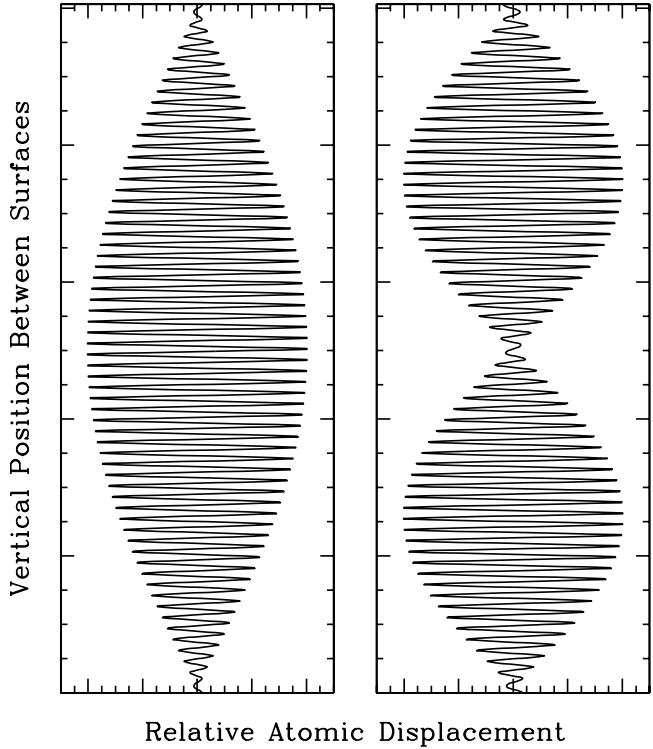}}
\caption{
Atomic displacement pattern for the first two modes below the band gap for a 
2,000-layer, bare (6,2) superlattice substrate.  The vertical axis denotes the 
position of a layer of atoms relative to the surface, and the horizontal axis
denotes the relative displacement of each layer away from equilibrium.  The top
and bottom of the vertical axis are the positions of the top and bottom surface
of the substrate.
}
\label{beat}
\end{figure}

To understand the source of this weaker coupling, we investigate the
nature of the bulk modes in the enhanced spectral region below the band
gap.  Figure \ref{beat} illustrates the atomic displacement pattern for
the first two modes below the band gap for a 2,000-layer, bare (6,2)
superlattice substrate.  The vertical axis again denotes the position 
of a layer of atoms relative to the surface, but in this figure, unlike 
Fig.~\ref{eigvec}, all substrate layers are shown.  These modes have a
beat-like pattern, with the comparatively short-wavelength oscillations
spanning about 16 ({\it i.e.,} $2N$) atomic layers per cycle and the 
long-wavelength oscillations exhibiting fixed-end standing-wave character.

The existence of these beat-like modes is tied to the formation of the
band gap.  For wavelengths around $2N$ atomic layers in a period-$N$ 
superlattice, it becomes possible to form linear combinations of 
homogeneous-substrate modes that have maximal atomic displacement solely 
in either material-$A$ or material-$B$ regions of the superlattice.  
These linear combinations are stabilized by the coherence with the 
superlattice modulation, and they give rise to a band gap because of the 
inverse-square-root dependence of the frequency on the mass.  The beat
patterns form because the linear combinations superpose modes with 
similar spatial frequencies ({\it i.e.,} wavelengths).  There are many 
different ways to form linear combinations that are commensurate with 
the superlattice modulation, each one distinguished by the number of 
nodes in the resulting envelope waveform.  This is the source of the
spectral enhancement below and above the band gap.

For the (6,2) system, the adsorbate FT motion is in resonance with
these beat-like modes.  However, since these modes have very small
amplitude at the surface of the superlattice, the coupling between
them and the FT mode is very weak.  Therefore, despite the large
density of bulk modes available for coupling, the resulting 
resonance peak remains very narrow, and the FT-mode remains 
long-lived.

\begin{figure}
\epsfysize=3.00in
\centerline{\epsfbox[18 324 592 718]{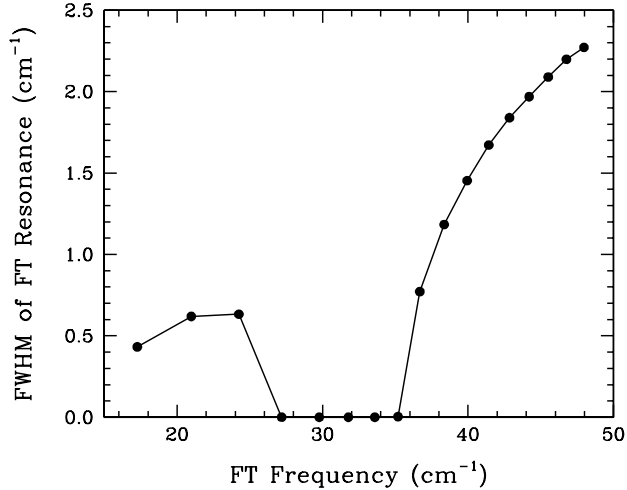}}
\caption{
Dependence of the FT resonance width on the FT frequency for the (6,2) 
superlattice.
}
\label{freqwid}
\end{figure}

To elucidate more fully the effect of the envelope function of the
beat-like modes on FT vibrational relaxation, we examine how the 
width of the FT resonance varies with FT frequency, for a given 
superlattice substrate.  We are able to accomplish this by varying 
the value of the force constant, $K_{as}$, coupling the adsorbates 
to the substrate,\cite{sumrule} since the FT frequency is a smooth, 
monotonically increasing function of $K_{as}$.  Figure \ref{freqwid} 
displays the computed dependence of the resonance width on the FT 
frequency for the (6,2) superlattice.\cite{swap}  As expected, 
the resonance width is zero for FT frequencies in the 
superlattice-induced bulk band gap.  The weak coupling of the FT
to the beat-like modes near the gap edges is seen as a rapid 
decrease in the resonance width when approaching the band-gap
region from either side.  Away from the bulk band gap and its
adjacent spectrally enhanced regions, the resonance width is a
monotonic, more gently increasing function of the FT frequency.

One concern regarding the analysis in this paper is that 
superlattices with very short modulation periods are required.  
As a result, the interfaces between the $A$ and $B$ layers must be 
sharp on the scale of one or two atomic layers.  This sharpness may 
be difficult to achieve for metallic superlattices due to interfacial 
diffusion or miscibility.  A potential solution to this concern would 
be to use a semiconductor superlattice for the substrate.  These 
materials can be grown with very sharp interfaces, and therefore may 
accommodate the short modulation periods that are required for the 
application described in this paper.  In order to restore the chemical 
environment of the adsorbate-metal interface, one could first deposit 
several layers of metal ({\it e.g.,} copper) on the surface of the 
superlattice, and then form the molecular overlayer on the supported 
metal film.

\section{Conclusions}

One of the most important goals in surface science is to be able
to control dynamical processes taking place at a surface.  We have
proposed substrate patterning as a means to control the relaxation
dynamics of low-frequency frustrated-translational vibrations of an 
adsorbate overlayer.  This proposal emerges from our recent work 
demonstrating that the dominant relaxation mechanism for these modes 
is resonant emission of phonons into the substrate.  Specifically, we 
have considered short-period superlattices as candidate substrates,
because the mass modulation of a superlattice causes gaps to open in
the phonon density of states.  Varying the superlattice composition
allows one to control the size and placement of these band gaps,
thereby controllably modifying the coupling between the adsorbate 
overlayer and the substrate.  We view the successful experimental
demonstration of the behavior predicted here as a stringent test of
our theory of resonant adsorbate-substrate coupling.  Moreover it
would establish substrate patterning as an important tool in 
surface dynamical research.

\section*{Acknowledgments}
Financial support for this research was provided by the National 
Science Foundation under Grant No.~DMR 97-02514, the Air Force 
Office of Scientific Research, Air Force Materiel Command, USAF, under 
Grant No.~F49620-00-1-0170, and the University of Georgia Research
Foundation.  SPL acknowledges support of the Donors of The Petroleum 
Research Fund, administered by the American Chemical Society.  AMR 
acknowledges support of the Alfred P. Sloan Foundation.


\begin{thebibliography}{MM}
\bibitem{swap} Figure \ref{freqwid} corresponds to the case where the 
superlattice substrate terminates on the more massive material, $B$.  
This choice has the effect of suppressing the intra-gap surface state.  
The alternate choice of terminating on an $A$-type layer leads to a more 
complicated picture involving an avoided crossing of the FT and surface 
states as a function of $K_{as}$.
\end{thebibliography}
\end{document}